\documentclass{ws-p8-50x6-00}

\begin{document}

\title{Angular Distributions in Higgs Decays}

\author{A. Mondragon and R. E. Allen}

\address{Department of Physics, Texas A\&M University, College Station,
Texas 77843, USA\\ e-mail: mondragon@tamu.edu}

\maketitle

\abstracts{A new theory yields unconventional results for the angular
distribution of products like $b-\bar{b}$ jets in Higgs decays.}

According to a new fundamental theory,\cite{allen1,allen2} Higgs
fields have an unconventional equation of motion
\begin{equation}
-\left( g^{\mu \nu }D_{\mu }D_{\nu }+i\overline{m}e_{\alpha }^{\mu }\sigma
^{\alpha }D_{\mu }\right) \Phi _{h}-\mu _{h}^{2}\Phi _{h}+\overline{b}\left(
\Phi _{h}^{\dagger }\Phi _{h}\right) \Phi _{h}=0.
\end{equation}
(See (13) of Ref. 2.) At very high energies and field strengths,
this goes over to the usual form
\begin{equation}
\left[ -g^{\mu \nu }D_{\mu }D_{\nu }-\mu _{h}^{2}+\overline{b}\left( \Phi
_{h}^{\dagger }\Phi _{h}\right) \right] \Phi _{h}=0.
\end{equation}
At low energies and field strengths, however, it becomes
\begin{equation}
\left[ ie_{\alpha }^{\mu }\sigma ^{\alpha }D_{\mu }+\widetilde{\mu }-
\widetilde{b}\left( \Phi _{h}^{\dagger }\Phi _{h}\right) \right] \Phi _{h}=0
\end{equation}
where $\widetilde{\mu }=\mu _{h}^{2}/\overline{m}$ and $\widetilde{b}=
\overline{b}/\overline{m}$.

\bigskip
In flat spacetime $\left(e_{\alpha }^{\mu }=\delta _{\alpha }^{\mu }\right)$,
and with gauge couplings ignored, (3) reduces to
\begin{equation}
i\sigma ^{k}\partial _{k}\Phi _{h}=\left[ -i\partial _{0}-\widetilde{\mu }+%
\widetilde{b}\left( \Phi _{h}^{\dagger }\Phi _{h}\right)
\right] \Phi _{h}
\end{equation}
for the original Higgs fields. For the physical Higgs boson $\phi
_{h}$, on the other hand, a standard treatment gives the above equation of
motion with $-\widetilde{\mu }\to \bar{\mu }=2\widetilde{\mu }$
and different self-interaction terms.\cite{cheng}
If these self-interactions are also neglected, we obtain
\begin{equation}
-\partial ^{k}\partial _{k}\phi _{h}=\left( i\partial _{0}-
\bar{\mu }\right) ^{2}\phi _{h}
\end{equation}
so that
\begin{equation}
p^{2}=\left( \omega -\bar{\mu }\right) ^{2}
\end{equation}
and
\begin{equation}
\omega =p+\bar{\mu }
\end{equation}
for the physical branch, where $p$ is the magnitude of the
3-momentum. Because the present picture
involves (physically acceptable) violations of Lorentz
invariance,\cite{allen1,allen2,allen3} the
energy-momentum relation of the Higgs is quite different from
\begin{equation}
\omega =\sqrt{p^{2}+m_{h}^{2}}
\end{equation}
and observable properties like angular distributions in decays are
consequently predicted to be different.

\bigskip
A simple illustration is the kinematics of the $h^{0} \to q \bar{q}$
decay, with $q$ representing a light particle like the bottom
quark.\cite{gunion,quigg} Conservation of energy implies that
\begin{equation}
p_{h}+m_{h}=p_{q}+p_{\bar{q}}
\end{equation}
or
\begin{equation}
p_{h}^{2}=p_{q}^{2}+p_{\bar{q}}^{2}+2p_{q}p_{\bar{q}}+m_{h}^{2}-
2m_{h}(p_{q}+p_{\bar{q}})
\end{equation}
if the mass of $q$ is neglected. (We now write $m_{h}$ for $\bar{\mu }$.)
Conservation of 3-momentum gives
\begin{equation}
p_{h}^{2}=p_{q}^{2}+p_{\bar{q}}^{2}+2\,{\bf p}_{q}
{\bf\cdot }{\bf p}_{\bar{q}}
\end{equation}
or, in terms of the opening angle $\theta$,
\begin{equation}
p_{h}^{2}=p_{q}^{2}+p_{\bar{q}}^{2}+2\,p_{q}p_{\bar{q}}\cos \theta.
\end{equation}

\bigskip
The predicted angular distribution
\begin{equation}
\cos \theta =1+\frac{m_{h}^{2}}{2p_{q} p_{\bar{q}}}-
m_{h}{\frac{p_{q}+p_{\bar{q}}}{p_{q} p_{\bar{q}}}}
\end{equation}
is then quite different from the standard result
\begin{equation}
\cos \theta =1-{\frac{m_{h}^{2}}{2p_{q}p_{\bar{q}}}}
\end{equation}
for energies significantly above threshold. For example, if the
emerging particles or jets have equal energy, and the total energy is twice
the threshold for Higgs production, the opening angle is 120$^{\circ}$ in the
present picture, rather than 60$^{\circ}$.

\section*{Acknowledgement}
This work was supported by the Robert A. Welch Foundation.

\end{document}